# Grip force as a functional window to somatosensory cognition

Birgitta Dresp-Langley

Centre National de la Recherche Scientifique CNRS UMR 7357

<u>birgitta.dresp@cnrs.fr</u>

## Abstract

Analysis of grip force signals tailored to hand and finger movement evolution and changes in grip force control during task execution provide unprecedented functional insight into somatosensory cognition. Somatosensory cognition is the basis of our ability to act upon and to transform the physical world around us, to recognize objects on the basis of touch alone, and to grasp them with the right amount of force for lifting and manipulating them. Recent technology has permitted the wireless  monitoring of grip force signals recorded from biosensors in the palm of the human hand to track and trace human grip forces deployed in cognitive tasks executed under conditions of variable sensory (visual, auditory) input. Non-invasive multi-finger grip force sensor technology can be exploited to explore functional interactions between somatosensory brain mechanisms and motor control, in particular during learning a cognitive task where the planning and strategic execution of hand movements is essential. Sensorial and cognitive processes underlying manual skills and/or hand-specific (dominant *versus* non-dominant hand) behaviors can be studied in a variety of contexts by probing selected measurement *loci* in the fingers and palm of the human hand. Thousands of sensor data recorded from multiple spatial locations can be approached statistically to breathe functional sense into the forces measured under specific task constraints. Grip force patterns in individual performance profiling may reveal the evolution of grip force control as a direct result of cognitive changes during task learning.  Grip forces can be functionally mapped to from-global-to-local coding principles in brain networks governing somatosensory processes for motor control  in cognitive tasks leading to a specific task expertise or skill. Under the light of a comprehensive overview of recent discoveries into the functional significance of human grip force variations, perspectives for future studies in cognition, in particular the cognitive control of strategic and task relevant hand movements in complex real-world precision tasks, are pointed out.

## Introduction

Early in life we learn to anticipate the shape, size, weight, and feel of physical objects by touching and manipulating them with our two hands. Seeing these objects is helpful during this learning process, but it is by no means indispensable. A large amount of primate cognition depends, indeed, on somatosensory processing and learning (De Haan and Dijkerman, 2020), which provides the most important basis for cognition in the blind (Wittenberg *et al.*, 2004; Arioli, Ricciardi, and Cattaneo, 2021). Somatosensory cognition generates awareness of our body, its different parts and their function, and is critically important to almost every daily activity that involves motor planning and control. Somatosensory cognition is the basis of the human ability to act upon and to transform the physical world. Such ability includes recognizing physical objects on the basis of touch alone, detecting and anticipating an object's temperature or sharpness and the pain it may cause when we touch or grasp it. Gripping objects with the right amount of force for lifting them, anticipating the consequences, or our potential inability of doing so, is critical to our survival. In short, somatosensory cognition enables us to efficiently interact with complex environments in a variety of contexts.

## Neural circuitry

All cognition requires sensory experience for the formation of neuronal connections by self-organized learning (e.g. Singer, 1986). Neuronal activity and the development of functionally specific neural networks in the continuously learning brain are modulated by sensory signals processed in the somatosensory cortical networks, the so-called S1 map (Wilson and Moore, 2015). S1 refers to a neocortical area that responds primarily to tactile stimulations on the skin or hair, and plays a critical role in grip force control, in interaction with multiple sensory areas. Somatosensory neurons have the smallest receptive fields and receive the shortest-latency input from the receptor periphery, and their cortical functional organization is conceptualized in current state of the art (Wilson and Moore, 2015; Braun et al., 2001; Arber, 2012) in terms of a single map of the receptor periphery. The somatosensory cortical network has a modular functional architecture with highly specific connectivity patterns (Arber, 2012; Tripodi and Arber, 2012), binding functionally distinct neuronal subpopulations from other cortical areas involved in sensory processing into motor circuit modules at several hierarchical levels (Tripodi and Arber, 2012; Merel J, Botvinick M, Wayne, 2019). These functional modules display a hierarchy of interleaved circuits connecting via inter-neurons in the spinal cord, in visual, auditory and olfactive sensory areas, and in motor cortex, with feed-back loops and bilateral communication with the supraspinal centers (Arber, 2012; Tripodi and Arber, 2012; Mendoza and Merchant, 2014). Anatomically adjacent to motor cortex, S1 is functionally connected to all the other sensory areas (Mendoza and Merchant, 2014) involved in processing stimuli in the environment (Fig. 1). Somatosensory afferents reach the frontal lobe, feeding into circuitry for prefrontal responses to somatosensation and the conscious control of motor and grip force behavior (Romanski, 2012). The from-local-to-global functional organization of the motor brain displays specific connectivity patterns in large-scale functionally specialized networks the activities of which correlate with specific behaviors and motor control in interaction with sensory processing, in particular for limb movement control (Borra and Lupino, 2019). Current state of the art suggests a precisely timed neurogenesis for the developmental specification of the somatosensory brain (Weiss et al., 2000; Bassett et al., 2006). In non-human primates (Merzenich et al., 1984) and in human patients (Wall, Xu, and Wang), somatosensory representations of the fingers left intact after amputation of others on the same hand become expanded in less than ten days after amputation by comparison with representations in the intact hand. Such network expansion reflects functional resilience through spontaneous and self-organizing synaptic plasticity in the somatosensory brain and offers insight into the evolution of the human hand and manual abilities.

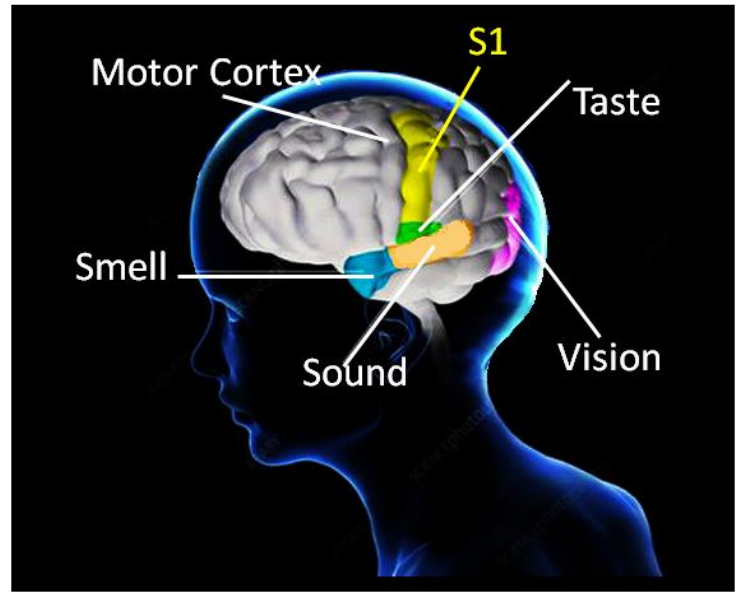

**Figure 1:** Anatomically adjacent to the motor cortex, the somatosensory brain map (S1) refers to a neocortical area that responds primarily to tactile stimulations on the skin and plays a critical role in grip force control through functional interactions with multiple sensory areas. Somatosensory afferents also reach the frontal lobe (not illustrated here), providing circuitry for prefrontal responses to somatosensation and the conscious control of grip force and motor behavior in general. Somatosensory cognition is a result of these complex functions, which develop and evolve with direct contact and interaction with physical environments.

**Synergy between sensory processing and cognitive control**

Manual ability and grip force behavior have evolved as a function of active constraints (Young, 2003) in harmony with other sensory systems such as somatosensation, vision, hearing, and smell (Reichert and Bolz, 2018). Thus, object-relevant cues provided by somatosensory, visual, auditory, and olfactory stimulation play an important role in grip force scaling. Somatosensory feedback is critical for processing the shape and texture of objects we grip (Flanagan, Bowman, and Johansson, 2006) and for developing effective motor control strategies. Under conditions of insufficent somatosensory feed-back, individuals use memory representations from previous experience, and patients with massive somatosensory loss can still scale and time grip forces. This permits adjusting them across different object handling tasks on the basis of memory-based, anticipatory, online, control processes to compensate for the loss of somatosensory feedback (Parry et al., 2021). The force with which we grasp objects is also determined by their visually perceived size, shape, and weight (Jenmalm and Johansson, 1997). Visual cues allow to anticipate object dynamics related to grip force required for a given task (Maiello et al., 2021), for example. Interactions between somatosensory and visual processing are reflected by differential modulation of finger grip forces when object shape and object density cues are congruent, and symmetrical force sharing when shape and density cues are incongruent (Lee-Miller et al., 2016). When objects are manipulated to have a different mass than would be expected for a given material on the basis of visual inspection, the initial grip forces deployed to lift them are systematically scaled to the expected, not the real weights of the objects. After a few lifts, individuals begin to scale their forces to the actual weight and disregard the misleading visual cues. However, the grip forces only match the visual perception of weight when all differences between visually predicted and experimentally manipulated ("false") object weights are removed (Buckingham, Cant, and Goodale, 2009). This reveals a strong functional link between visual and somatosensory cognition in sighted individuals, who have learnt to rely on visual input for motor planning and control. Sounds can also be relevant to deploying grip forces necessary to lift masses, as evidenced by scaling of grip rates as a function of mass during auditory cueing, for example (Gonzalez, Dubrowski and Canahan, 2010). Sensations of disgust triggered by stimuli in the olfactory environment modulate neural activity in the somatosensory brain, and may prevent us from picking up objects, or make us grasp them with lesser force (Croy et al., 2016; Yang and Wang, 2020). However, while sensory cues may allow to anticipate grip forces required for a specific task, this changes when such cues are suddenly no longer available, or the subject is confronted with uncertainty about digit-force related object properties. Sensorimotor memories (Cole, Potash, and Peterson, 2008) form during life-long brain learning on the basis of previous grasp experience to permit adaptive, cognitively controlled grip force planning when physical object properties and behavioral consequences cannot be reliably sensed (Lukos, Choi, and Santello, 2013). Exploring grip forces deployed for grasping, lifting and manipulating objects under conditions of variable external constraints and sensory input reveals the dynamic functional changes taking place in directly measurable behavior.

**From anthropometric factors to cognitive fitness**

Prehensile force adjustment is an essential feature of tool use and object manipulation in daily life involving both sensori-motor and cognitive skills. Effects of anthropometric parameters such as global hand size or proportional size ratios on individual finger grip forces are subtle and generally minor in normal adults, while there appears to be a significant positive correlation between handgrip strength and arm-hand dimensions in athletes (Fallahi and Jadidian, 2011), suggesting that grip force profiling could be used for initial talent identification in handgrip-related sports, or in clinical settings to set specific training or performance rehabilitation benchmarks. However, sensorimotor memories contribute to the scaling of grip forces in all motor tasks (Cole, Potash, and Peterson, 2008), and control processes may kick in to compensate for anthropometric limitations. Grip force may thus not predict the future performance of beginners. Also, while grip force may initially be stronger in the dominant hand, this can reverse spontaneously during motor learning, depending on the necessity of a given task and environmental constraints (Bohannon, 2003; Li and Yu, 2011; Cai et al., 2018). In older adults (Rijk et al., 2016; Lee et al., 2022), lower grip forces have been associated with symptoms of cognitive dysfunction, mild cognitive impairment, or the incidence of dementia (Kobayashi-Cuya et al., 2018; Cui et al., 2021; Zammit et al., 2021; Kunutsor et al., 2022). Decline in multi-finger grip strength maybe thus be indicative of a decline in cognitive function rather than in muscular fitness *per se*. Stronger grip forces have been associated with better cognitive performance in 527 patients between 45 and 62 years of age diagnosed with major depression (Firth, Smith, and Sarris, 2020). Multi-finger grip force deployment involves sophisticated neural control mechanisms (Kilgour, Todd, and Starr, 2014), and it has been suggested that decline in multi-finger grip strength is a marker of brain health (Carson and Holton, 2022). In healthy subjects (Cai et al. , 2018) grip force is accurately scaled to just a little stronger than the minimum necessary to prevent the object from slipping out of the hand as a result of adaptive processes in finger force control during motor learning (Zatsiorsky and Latash, 2008; Cole, Potash, and Peterson, 2008). Impaired finger force control produces inefficient grip force scaling and a deficient temporal synchronization of grip and load force profiles is characteristic of Parkinson's disease (Fellows and Noth, 2004; Nowak and Hermsdörfer, 2006), Tourette's syndrome (Nowak et al., 2005), and pathologies of the cerebellum (Rost et al., 2005). Weaker grip strength has been related to slow information processing times and impaired executive functioning in ageing adults (Kobayahsi-Cuya et al., 2018). The predictive link identified between grip strength, mortality, future function, bone mineral density, fractures, cognition, depression, and problems associated with hospitalization (Bohannon, 2019) lead to recommend using grip strength as a stand-alone measurement, or as a component of a small set of measurements, for identifying older adults at risk of poor health. Thus, grip force profiling could prove a useful tool for the objective evaluation of treatment approaches to cognitive and neurological disorders (Lee et al., 2022), and offer novel insights into physiological and cognitive dynamics of ageing.

**Finger-specific effects**

The contribution of each finger to overall grip strength, coarse adjustments, and finer grip force control vary across cognitive tasks and their requirements for motor planning and execution. While the middle finger is critical for lifting and manipulating heavy objects in three dimensions (Kinoshita, Kawai, and Ikuta, 1995), the ring finger and the small finger mostly control subtle grip force modulation (Wu, Zatsiorsky, and Latash, 2002; Cha et al., 2014; Batmaz et al., 2017; De Mathelin et al., 2019; Liu et al., 2020; Dresp-Langley et al. , 2020; Liu and Dresp-Langley, 2021) necessary for precision tasks such as surgery. In precision tasks, the contribution of the index finger to total grip force is often the smallest, and there seem to be no significant differences between men and women in this regard (e.g. Kinoshita, Kawai, and Ikuta, 1995; de Mathelin et al.; 2019; Dresp-Langley et al., 2020). The amount of force applied by each digit depends on  various factors, however, including where the digits are placed when grasping (Fu and Santello, 2014). In some tasks the thumb may exert the largest forces during manipulation closely followed by the little finger (Naceri et al., 2017). In the absence of external constraints, the complex anatomy of the human hand allows for a large number of postures and force combinations to attain stable grips, and these functional synergies permit solving the problem of motor redundancy (Latash and Zatsiorsky, 2009; Fu and Santello, 2014; Naceri et al., 2017; Dresp-Langley et al., 2020). Multi-finger grip force control thus relies on self-organizing prehensile synergies  involving from-local-to-global functional interactions at several hierarchical stages from hand to brain and back (Zatsiorsky and Latash, 2008; Fu and Santello, 2014; Dresp-Langley, 2020a, b). The adaptive generation of both magnitude and rate of hand or finger forces are skill-specific and involve high-level cognitive control mechanisms in experts (e.g. Johansson and Cole, 1992; Eliasson et al., 1995). In pianists, for example, skilled finger force deployment and control has been directly related to musical expertise (Oku and Furuya, 2017). Skilled surgeons deploy grip forces more parsimoniously than novices (Judkins et al. , 2008), and their spatiotemporal grip force profiles show specific patterns characteristic of expertise by comparison with the profiles of novices or trainee surgeons (Abiri et al., 2019; Dresp-Langley et al. , 2020; Liu et al. , 2020).

**Functional aspects of robot-assisted motor tasks**

Under the light of the analysis here above, we define somatosensory cognition in terms of multi-sensorial synergies controlled by cognitive processes beyond sensation and perception. Cognitive control processes include perception-based decision, learning, skill, and experience (memory) during manual interaction with the environment (Romo and Lafuente, 2013). Grip force deployment can be exploited as a behavioral marker of the complex processes that consist of learning new manual and cognitive skills that involve finely tuned hand and finger movement control. Robot-assisted surgical training illustrates some of the perspectives offered by modern grip force sensor technology for the study of functionally significant cognitive changes during such novel motor task skill learning. Minimally invasive robotic surgery is an image-guided, high-precision task, where the loss of haptic force feedback spontaneously yields stronger hand grip forces during task execution, especially in novices (Abiri et al., 2019). This can result in unnecessary or excessive tissue damage in a patient, and novices therefore have to learn to scale their finger forces accordingly. Also, robotic surgery systems require man-machine interaction under conditions of limited degrees of freedom for hand and finger movements to manipulate the surgical tools attached to the system in any direction that may be necessary for a given intervention. This represents a considerable constraint for motor planning and control to which a novice needs to adapt to by learning. In addition, the novice has to process critical information about what his/her hands are doing in a real-world environment while looking at a two-dimensional or virtual three-dimensional representation of that environment displayed on a monitor. This virtual information needs to be correctly interpreted by the brain to ensure a safe and effective intervention. Veridical information about real-world depth is missing from the image representations, and instead of looking down on his/her hands, the surgeon often only sees the tool-tips attached to the system. Due to a variety of camera and image calibration problems, the tool movements displayed on the screen may not match the real-world movements in time and space. The loss of real-world depth input and veridical space scale information significantly affects the performance of novices, and adapting to these constraints is only possible through a long period of training (Batmaz et al., 2016, 2017a,b, 2018). In short, by profoundly challenging solidly formed perceptual, somatosensory, and cognitive representations of space, scale, and relative distance for manual coordination, robot-assisted precision tasks constitute a fertile study ground for adaptive mechanisms in human spatial cognition and hand and finger movement calibration for effective action. Proficiency and expertise in the control of a robotic surgery system is reflected by lesser grip force deployment by the middle finger during task execution accompanied by shorter task execution times (Figure 2). Grip force profiling permits tracking the evolution of a surgeon's individual motor behaviour during skill training and offers insight into cognitive and functional changes taking place as performance evolves with training and specific sensory feed-back. Haptic cues (Abiri et al., 2019), sound feed-back, or visual cueing by augmented reality can promise effective for modulating excessive grip forces when visual input insufficient (Dresp-Langley, 2020). Studying the interaction between grip force evolution, sensory feed-back, and the convergence of speed-precision trade-offs (Dresp-Langley, 2018) during motor relearning in such novel task scenarios offers further perspectives into functional changes. Such grip force studies could be combined with EEG recordings to highlight features that represent dynamic changes in the functional brain network (Shafiei, 2021) correlated with a given performance or learning curve.

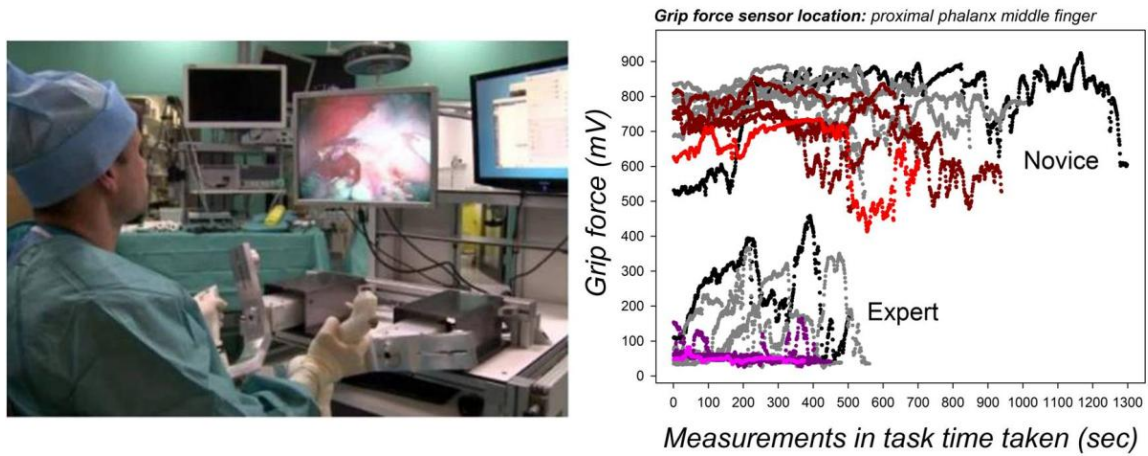

**Figure 2:** Distinct spatio-temporal grip force patterns (on the right) tell the skilled expert from the novice at a glance in 2D image-guided robot-assisted (left) simulations in a 3D surgical task space (de Mathelin et al., 2019; Dresp-Langley et al., 2020). The ten graphs plotted for each individual represent grip forces sampled every 20 milliseconds in ten successive training sessions from these studies. The colored graphs indicate the individual grip forces from the last session. Human-robot interaction represents a task context that profoundly challenges solidly formed perceptual, somatosensory, and cognitive representations of space, scale, and relative distance for sensori-motor coordination.

Comparing grip force profiles corresponding to thousands of individual measurements, collected from specific sensor positions on anatomically relevant parts of the finger and hand regions of the dominant and non-dominant hands of highly skilled experts and beginners, has enabled the discovery of distinct functional correlations and spatio-temporal patterns that tell the expert from the novice at a glance. A wireless sensor glove hardware-software system, specifically designed for such studies and described in detail elsewhere (e.g. de Mathelin et al., 2019) has generated these novel insights. The complex prehensile control strategies underlying skill formation in perceptual and motor relearning in robotic surgery are reflected by distinct adaptive grip force changes translating functional plasticity of somatosensory representations in the learning novice (Liu and Dresp-Langley, 2021). With the development of robot-assistance almost everywhere, from the workplace to the private home, such situations are likely to become more and more common.

**Discussion**

Hand and finger grip forces deployed for grasping, lifting, controlling, and manipulating objects in various tasks involving hand-object movements in three dimensions permit studying somatosensory cognition. Finger grip strength and manual force deployment reveal functional links between somatosensory mechanisms for motor control, and the sensorial processing of sound intensities and visual object properties. Thus, grip force analyses offer multiple perspectives for studying cognitive processes related to hearing or visual cognition. Multi-finger grip force deployment involves sophisticated neural control mechanisms. Its analysis as a marker of brain health offers new perspectives for clinical studies of cognitive disorders and/or decline. Recently developed wireless sensor technology enables the continuous and non-invasive monitoring of an individual's manual grip behaviour in real time. Individual grip force measures directly translate into spatiotemporal grip force profiles for different locations on the fingers and/or palm of the hand, and offer perspectives for studying how the human perceptual and cognitive system adapts to novel situations that challenge our solidly formed perceptual, somatosensory, and cognitive representations of scale and space. Wearable sensor technology offers innovative solutions in sensor design, electronics, and signal processing for statistical analysis of thousands of individual data. In addition to revealing functional aspects of cognitive control for action during task skill evolution, real-time grip force sensing by wearable systems can also help prevent incidents, because it includes the possibility of sending a signal (sound or light) to the operator whenever his/her grip force exceeds a critical limit before the damage is done. Although a growing amount of research is currently devoted to this area, new experimental design approaches, methods for data analysis, visualization, and knowledge representation are needed. These aspects represent an open challenge for the scientific community in this still novel field.